\documentstyle[prl,epsfig,aps]{revtex} 
\begin{document}
\draft

\title{The order-disorder transition of the $(3 \times 3)$Sn/Ge(111) phase}

\author{
L. Floreano,$^{1,*}$
D. Cvetko,$^{1,2}$
G. Bavdek,$^{1,2}$
M. Benes,$^{1,3}$
and A. Morgante$^{1,3}$
}

\address{
$^{1}$ Laboratorio TASC, Istituto Nazionale per la Fisica della Materia, 
Basovizza SS14 Km 163.5, I-34012 Trieste, Italy}

\address{ $^{2}$ Jo{\v z}ef Stefan Institute, Department of Physics, 
Ljubljana University, Ljubljana, Slovenia}

\address{
$^{3}$ Department of Physics, University of Trieste, Via 
Valerio 2, I-34100 Trieste, Italy}

\narrowtext
\onecolumn

\maketitle
\begin{abstract}
We have measured the long range order of the $\alpha$-phase of Sn on the Ge(111) 
surface throughout the $(3 \times 3) \rightarrow 
(\sqrt 3 \times \sqrt 3)$R30$^{\circ}$ phase transition. 
The transition has been found of the order-disorder type with
a critical temperature T$_{c} \sim 220$~K. The expected 3-State 
Potts critical exponents are shown to be consistent with the 
observed power law 
dependence of the $(3 \times 3)$ order parameter and its
correlation length close to T$_{c}$, thus excluding a 
charge density wave driven phase transition.

\end{abstract}
\vspace{1cm}


\narrowtext
\twocolumn

Growing attention has been drawn in the past years to the $\alpha$-phase 
(1/3 monolayer) of both Pb and Sn on Ge(111), which undergoes a 
transition from the room 
temperature, RT, $(\sqrt 3 \times \sqrt 3)$R30$^{\circ}$ phase to the 
low temperature, LT, $(3 \times 3)$ one. On the basis of scanning 
tunnelling microscopy (STM) experiments, this transition was claimed to 
be the manifestation of a surface charge density wave (SCDW), i.e. a 
periodic redistribution of charge, possibly accompanied by a small 
periodic lattice distortion, which determines a change of the 
surface symmetry \cite{carpinelli,carpinelli2,scandolo,gonzalez}.
This model interpretation would imply relevant effects of electronic 
nature, such as magnetism of the LT phase \cite{scandolo} and 
electron correlation effects \cite{goldoni}. 
As further experiments with different techniques were being 
performed to explore the properties of this system, 
increasing doubts were cast about the SCDW model. The 
research was then focused on the determination of the atomic structure of 
the RT and LT phases, in an effort to discriminate between 
different transition models, as well as on the study of the role played by 
defects, which were seen to stabilize the LT 
phase \cite{weitering,melechko} and to determine a metal to 
semiconductor transition \cite{kidd}. 

At present, the atomic structure of the LT $(3 \times 3)$Sn/Ge(111) phase has 
been determined with a substantial agreement between X-ray 
diffraction (XRD)\cite{bunk,zhang} and photoelectron 
diffraction\cite{floreano} measurements. 
The Sn atoms occupy the T$_{4}$ sites above the Ge(111) lattice, but 
one Sn atom, out of three, per unit cell is vertically displaced by 
$\sim 0.3$~\AA~ in the outward direction. This distortion 
strongly affects the substrate lattice too, where the three nearest 
neighbour Ge atoms are found to follow the Sn atom in its vertical 
displacement. The most recent calculations in local density 
approximation reproduce quite exactly this vertically rippled 
structure\cite{degironcoli,ortega}.

The RT phase structure is a more controversial issue. 
The analysis of the most recently published 
XRD measurements slightly favours a structure where the vertical 
ripple disappears and the Sn atoms occupy equivalent T$_{4}$ sites at 
the same height level\cite{bunk}. On the basis of this observation and taking 
into account the strong distortion of the substrate too, a short 
interaction range displacive phase transition (pseudo Jahn-Teller distortion) 
appeared to describe the system 
behaviour more adequately than a long interaction range SCDW\cite{pick}.
On the other hand, the spectroscopic measurements give a result 
that seems incompatible with both the STM and XRD measurements. 
The Sn 4d core level spectrum associated with the $\alpha$-phase can 
be fitted with two components (with intensity ratio of 1:2), which 
have been attributed to the two types of Sn atoms in the LT $(3 \times 
3)$ phase \cite{floreano,lelay,uhrberg,avila99}. The same two components are 
found for both the LT and RT phases, i.e. the vertical ripple is 
expected to remain also at the RT $(\sqrt 3 \times \sqrt 3)$R30$^{\circ}$ 
phase \cite{lelay,uhrberg,avila,uhrberg2}. This result clearly points 
to an order-disorder character of the transition. In addition, 
a model based on molecular dynamics calculations has shown that the 
Sn atoms can jump between the two height levels,
 but the ratio 
between the number of up and down atoms remains equal to 
1:2, even above the transition temperature, where the jumping rate 
would be beyond the STM time resolution\cite{avila}. 

Most strikingly, all the experimental efforts for discriminating among 
the possible transition models have been devoted to the determination 
of the local atomic structure of the two Sn phases. To our knowledge, 
the diffraction techniques (both electrons and X-rays) have not been 
employed to quantitatively study the temperature behaviour of the diffraction pattern, 
i.e. of the surface order parameter which describes the thermodynamics of the 
whole system. We have thus performed He atom scattering (HAS) 
experiments to directly measure the surface order parameter, 
thus characterizing the phase 
transition. This experimental technique
is only sensitive to the surface charge density (like STM), 
in addition it is a long range order probe and offers the 
advantage of a short interaction time (10$^{-13}$ sec).

The experiment has been performed with a compact HAS apparatus which 
is described in more detail elsewhere\cite{apparato}. 
A 20.1~meV He 
beam ($k=6.3$~\AA$^{-1}$) has been used for the present 
measurements; the corresponding instrumental resolution and angular reproducibility yield
a transfer width exceeding 1200~\AA.
The measurements have been performed on different samples from 
the same Ge(111) wafer.
The surface has been cleaned by 1~keV Ar$^{+}$ ion bombardment and 
annealing up to 1100~K, thus obtaining a good $c(2\times 8)$
pattern. 
Sn has been evaporated at T$_{cell}$~=~1220~K from a Knudsen cell 
hosted into a liquid nitrogen cryopanel. The base pressure never 
exceeded $1 \cdot 10^{-9}$~mbar during evaporation and remained at
$2 \cdot 10^{-10}$~mbar during the measurements.

\begin{figure}
\includegraphics[width=.46\textwidth]{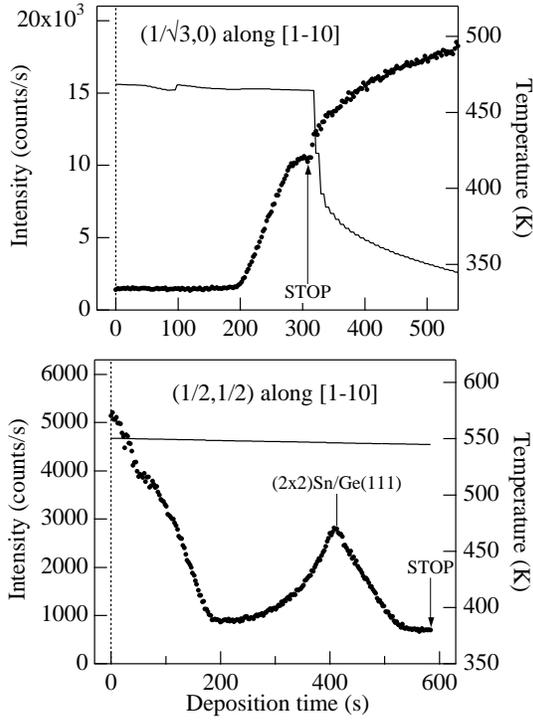}
\caption{Upper panel:
 intensity of the ($1/\sqrt 3$,0) peak taken along the 
$[1\overline{1}0]$ direction during Sn deposition at an approximate 
rate of 15~min/ML (filled markers, left axis). The 
surface temperature is also reported (full line, right axis).
Lower panel:  intensity of the (1/2,1/2) peak (with labelling relating to the 
 $c(2 \times 8)$ clean surface phase) taken along the 
$[1\overline{1}0]$ direction during Sn deposition 
 at an approximate rate of 30~min/ML (filled markers, left axis). The 
surface temperature is also reported (full line, right axis).
The occurence of the intermediate $(2 \times 2)$ Sn phase is also 
labelled.
}
\label{deposition}
\end{figure}

\begin{figure}
\includegraphics[width=.46\textwidth]{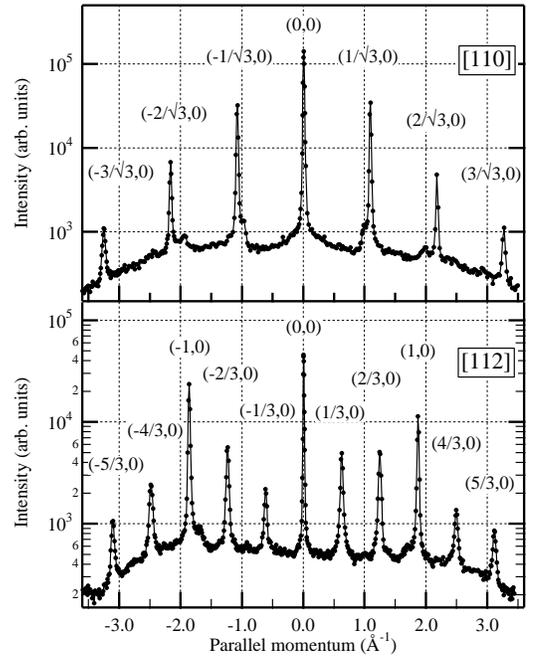}
\caption{
He diffraction patterns 
(k$_{He}$ = 6.3 \AA$^{-1}$) taken from the 
$\alpha$-phase of Sn on Ge(111) at 140~K. 
Upper panel: measurement along the  $[1\overline{1}0]$ direction, 
the intensity is on a logarithmic scale.
The fractional order peaks can be equivalently labelled as 
$(\pm \frac{n}{\sqrt 3},0)$, relating to the 
$(\sqrt 3 \times \sqrt 3)$R30$^{\circ}$ phase, 
or $(\pm \frac{n}{3},\pm \frac{n}{3})$, 
when relating to the $(3 \times 3)$ phase.
Lower panel: measurement along the [112] direction.
 The fractional order peaks 
$(\pm \frac{n}{3},0)$ are characteristic of the $(3 \times 3)$ 
phase. A mean $(3 \times 3)$ domain size of 200~\AA~ is estimated 
from the width of the $(\pm \frac{n}{3},0)$ peaks.
}
\label{pattern}
\end{figure}

For the preparation of the $(3 \times 3)$ phase of Sn on Ge(111) we have 
followed a procedure slightly different from the standard one reported 
in the literature, where Sn is usually evaporated on samples held at RT
 and successively annealed to T$_{s} \sim$~500~K.  
 We have monitored the intensity of the diffraction peak (1/$\sqrt 
 3$,0) along the $[1\overline{1}0]$ surface direction 
while depositing at T$_{s} \sim$~500~K (see upper panel of 
Fig.~\ref{deposition}). 
 After 2/3 
of the total exposure used in our work ($\sim 10$~min), the
$(\sqrt 3 \times \sqrt 3)$ pattern starts to appear, but the 
deposition is stopped only at the maximum intensity of the (1/$\sqrt 
 3$,0) peak. The surface is then left to cool down. 
 Alternatively, one can monitor the half-integer peaks of the $c(2 \times 8)$ 
 pattern up to their disappearance, which only occurs 
 after the formation of the intermediate $(2 \times 2)$Sn/Ge(111) phase 
 \cite{floreano} (see lower panel of 
Fig.~\ref{deposition}). 
 The latter procedure is less accurate in the coverage calibration.
 The diffraction 
 patterns of the $(3 \times 3)$ phase taken at 140~K are shown in Fig.~\ref{pattern} 
 along both the $[112]$ and $[1\overline{1}0]$ directions. The diffraction peaks 
 along the $[112]$ are characteristic of the $(3 \times 3)$ phase, 
 while those along the $[1\overline{1}0]$ direction belong to the RT $(\sqrt 3 \times \sqrt 
 3)$R30$^{\circ}$ phase too. 

The full diffraction patterns for a few temperatures have been taken 
throughout the transition. Along the $[1\overline{1}0]$ direction, both the integer and 
fractional order peaks display the same behaviour, i.e. an exponential 
intensity decrease due to the Debye-Waller (DW) attenuation without any 
variation of their angular width.  It must be concluded that 
the $(\sqrt 3 \times \sqrt 
 3)$R30$^{\circ}$ long range order is maintained throughout the 
 130-300~K temperature range.

\begin{figure}
\includegraphics[width=.46\textwidth]{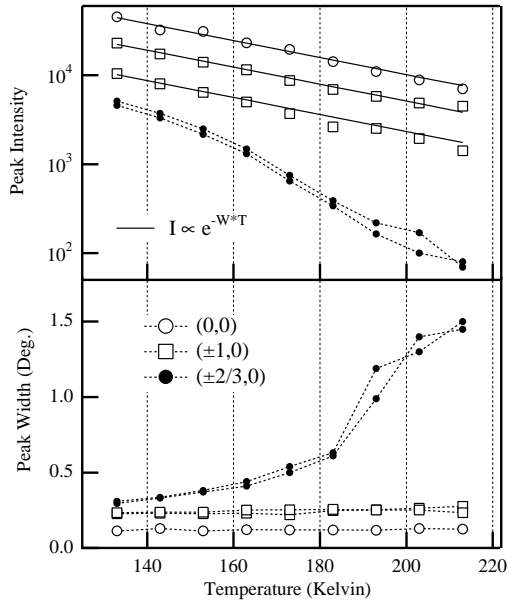}
\caption{ 
Temperature behaviour of the diffracted peak 
intensity and width (FWHM) taken along the [112] direction.
The integer order peaks (open markers) and the fractional ones 
(filled markers) have been taken from the same surface of Fig.~1, 
yielding a mean $(3 \times 3)$ domain size of 200~\AA.
Upper panel: the peak intensity is reported on a logarithmic scale to 
put in evidence the deviation from the Debye-Waller attenuation of the 
fractional order peaks. Full lines are best fit to the DW factor 
$e^{-W T}$ with $W = 0.022$~K$^{-1}$.
Lower panel: the diffraction peak FWHM is reported.
} 
\label{allpeaks}
\end{figure}

Much different is the diffraction pattern behaviour along the $[112]$ 
direction, as reported in Fig.~\ref{allpeaks}. In this case the 
specularly reflected peak and the integer peaks display the same 
exponential DW intensity decrease\cite{nota}, while the fractional order peaks 
present a steeper decrease with an inflection point at about 180~K. 
This observation is in contrast with the CDW model, whose predicted 
static distortion of the surface corrugation should strongly affect 
the He diffraction pattern. In particular, the expected metallicity of 
the RT phase\cite{scandolo,kidd} should flatten the surface charge 
density, thus leading to an increase of the He specular intensity (as 
observed for the $c(2 \times 8) \rightarrow (1 \times 1)$ transition 
on the Ge(111) surface\cite{meli}).
In addition, the full width at half maximum (FWHM) 
of the (0,0) and (0,$\pm$1) peaks 
remains unchanged, while the fractional order peaks display a strong 
broadening as the temperature increases. This latter indicates 
 a strong $(3 \times 3)$ domain wall proliferation, 
 thus suggesting the occurrence of an 
order-disorder phase transition \cite{nota2}.
In proximity of such an order-disorder equilibrium phase transition, all the
thermodynamic quantities scale as  power laws of the reduced temperature 
$t = \frac{T_{c}-T}{T_{c}}$. The corresponding critical exponents 
are only determined by the surface symmetry and  they 
indicate the universality class of the phase transition. 
According to the Landau's symmetry rules, 
this phase transition is expected to 
belong to the 3-state Potts universality class, due to the 
three-fold symmetry of the triangular surface lattice.

\begin{figure}
\includegraphics[width=.46\textwidth]{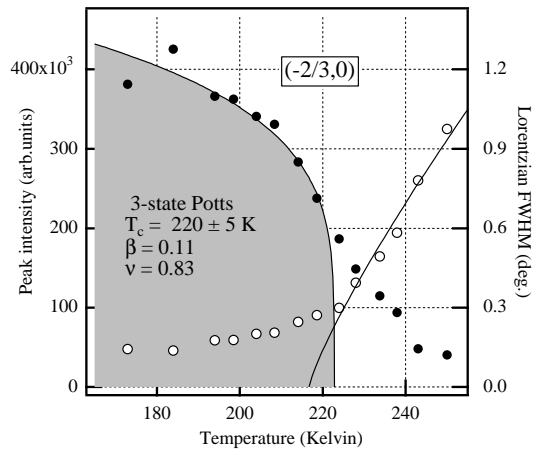}
\caption{ 
Temperature dependence of the (-2/3,0) He diffraction peak 
taken along the [112] direction from a surface with a $(3 \times 3)$ 
mean domain size of $\sim$~400~\AA. The peak intensity (filled 
circles, left axis) and the 
Lorentzian FWHM (open circles, right axis) were obtained by fitting 
the data to a
Voigt function. A Gaussian broadening of 
0.2$^{\circ}$ (FWHM), due to the 
instrumental resolution, has been used.
The peak intensity has been corrected for the DW 
factor, as obtained from Fig.~2.  
The full line is a fit of the Lorentzian 
widths to $|t|^{\nu}$ with the predicted 3-state Potts
 exponent $\nu = 
0.83$, thus yielding T$_{c}$~=~217~K. The shaded line 
represents the expected behaviour, $\beta = 0.11$, 
of the order parameter for the critical temperature of 223~K.
} 
\label{potts}
\end{figure}

To check the consistency of the order-disorder hypothesis 
we have followed in more detail and with larger statistics the temperature 
dependence of one of the diffraction peaks characteristic 
of the $(3 \times 3)$ long range order, which disappears above the 
critical temperature $T_{c}$. 
The temperature dependence of the diffracted peak has been found 
reversible, provided that the $(3 \times 3)$ domains extend at least 
a few hundreds of \AA ngstroms.
 The analysis of the temperature behaviour 
of the (-2/3,0) peak is reported in Fig.~\ref{potts}. 
In this case, the measurements were taken from a surface 
displaying a $(3 \times 3)$ mean domain size of 
400~\AA, i.e. twice that of Fig.~\ref{allpeaks}. 

In proximity of the transition temperature, 
the diffraction intensity for parallel momentum exchange {\bf K} 
 close to the reciprocal lattice vector {\bf G}=(-2/3,0) can be 
written as\cite{willis}: 

\begin{equation}
I({\bf K}, T) = \rho^2(T) \delta ({\bf K}-{\bf G}) + 
 \chi ({\bf K}-{\bf G}, T),
 \label{eq1}
\end{equation}

where $\rho$ stands for the order parameter and $\chi$ is the order 
parameter susceptibility, which accounts for the order fluctuations
close to $T_{c}$ (Fourier transform of the order parameter correlation 
function). 
 The order parameter vanishes at $T_{c}$ 
as $\rho \sim t^{\beta}$, and the fluctuations scale as 
 $\chi \sim |t|^{-\gamma}$. As a consequence, the order parameter susceptibility 
is the only contribution to the (-2/3,0) diffraction peak above the 
transition temperature and determines the peak shape. 
The susceptibility
 can be approximated to a Lorentzian and its width is proportional to the inverse of 
the $(3 \times 3)$
correlation length $\xi$, thus giving the spatial extent of the fluctuating domains. 
The susceptibility correlation length 
diverges at $T_{c}$ as $\xi \sim |t|^{-\nu}$.
The 3-state Potts critical exponents assume the 
fractional values $\beta = 1/9$, $\nu = 5/6$ and $\gamma = 13/9$ 
\cite{schick}.
 Experimentally, the 
susceptibility contribution is convoluted 
with the instrumental profile (which is assumed to be a 
Gaussian). In the experiment we have measured the (-2/3,0) peak 
profile at different temperatures and the analysis has been performed 
by fitting the data to a Voigt function with constant Gaussian width 
(corresponding to the Gaussian contribute obtained 
at the lowest achieved temperature). 
The resulting
Lorentzian width is shown in Fig.~\ref{potts} (open circles)
together with the (-2/3,0) peak intensity (filled circles).

The Lorentzian width displays a slight broadening below 220~K, 
possibly due to residual defects, while a much stronger broadening 
sets in above 220~K, when the order parameter intensity is strongly 
reduced. This observation points to an 
order-disorder phase transition, where the order parameter $\rho$ 
dominates the diffracted peak behaviour below $T_{c}$, according to 
eqn.~\ref{eq1}, and the Lorentzian should start to broaden after the 
disappearance of the term $I \propto t^{2\beta}$. For what 
concerns the critical exponents, we found a reasonable 
temperature range where the 
Lorentzian width can be fitted with the 
predicted $\nu = 5/6$ critical exponent, thus giving a transition 
temperature of 217~K. This transition temperature does not yield a satisfactory 
power law fit of the peak intensity. 
In this case, the expected $\beta = 1/9$ exponent 
is obtained for a slightly higher transition temperature. 
This discrepancy (yielding a critical temperature T$_{c} = 220 \pm 
5$~K with a corresponding dispersion of $\pm 20 \%$ in the 
critical exponents) is due to the rounding of both the order parameter and its 
correlation length close to T$_{c}$ and is to 
be related to the presence of defects within the $(3 \times 3)$ 
domains.

Both line and point defects produce a 
smearing of the equilibrium phase transitions\cite{sokolowski}. 
In fact, several kinds of point defects have been observed by STM on the 
$(3 \times 3)$Sn/Ge(111). Most of them are found to be
Ge substitutional 
impurities within the Sn overlayer, which are shown
to stabilize the LT phase in a local environment of the given 
$(3 \times 3)$ sublattice\cite{weitering}. 
This case resembles the order-disorder $c(4 \times 2) 
\rightarrow (2 \times 1)$ transition on the (001) surface of both Ge 
and Si. Due to defects, the expected 2D-Ising critical exponents 
have been only recently found for Ge(001)\cite{cvetko}, 
but are still lacking for Si(001)\cite{kubota}. 
This surface has been found to be strongly affected 
by point defects\cite{wolkow,tochihara}, 
which have been demonstrated to reduce the order parameter 
below $T_{c}$ and to smear out the 
transition \cite{inoue,nakamura}.  
A possible difference with the Sn/Ge system is the partial mobility 
of the Ge substitutional impurities
(they were seen to lie on a single $(3 \times 3)$
sublattice at 120~K, as opposed to the random distribution observed 
above 165~K)  \cite{melechko}. This observation led the authors 
of Refs.~\cite{weitering,melechko} to conclude that the transition 
is driven by a defect-defect interaction mediated by a 
SCDW, as a consequence the critical 
temperature was also predicted to decrease by decreasing the density of 
defects.
From He scattering, the individual role of each kind of defect cannot 
be discriminated, but it is well observed that a larger $(3 \times 3)$ 
domain size yields a higher critical temperature $T_{c}$ (compare the 
inflection points of Figs.~\ref{allpeaks} and \ref{potts}, taken 
from $(3 \times 3)$ domains yielding a mean size of 200 and 400~\AA, 
respectively). This observation is consistent with both the coverage 
dependence of $T_{c}$ 
for adsorbates order-disorder phase transitions (where it is demonstrated 
that the exact coverage, i.e. best surface quality, yields the highest 
critical temperature \cite{persson}) and the finite-size
scaling laws for order-disorder phase transitions, where a reduced 
domain size is usually observed to yield a lower estimate of 
Tc \cite{ferdinand,landau}.

On the basis of the present experiments, the SCDW phase transition must 
be excluded, since a clear order-disorder behaviour is displayed at 
the critical temperature. The occurrence of a late stage (higher 
temperature) displacive character of the
$(3 \times 3) \rightarrow (\sqrt 3 \times \sqrt 3)$R30$^{\circ}$ 
phase transition cannot be excluded {\it a priori} by the 
present study. 
In fact, all of the displacive phase transitions are expected to display an 
intermediate stage, where the order-disorder character
dominates\cite{bruce,tosatti}, even for a large temperature 
range \cite{toennies}. 
In the present case, an upper limit to the available temperature 
range is set to 550-600~K, where the $(\sqrt 3 \times \sqrt 3)$ 
phase is irreversibly decomposed into a new $(7 \times 7)$ phase. 
This temperature range is probably too small for 
displaying the onset of any displacive character.

M.B. acknowledges a grant by MURST cofin99 (Prot. 9902112831).

Added note: during the manuscript refereeing, a theoretical model 
suggesting a displacive transition was also proposed \cite{perez}, 
while new photoemission and photoelectron diffraction data 
excluded this hypothesis up to 500~K \cite{petaccia2}.


\begin{thebibliography}{99}
\bibitem{nota} in our experimental apparatus with fixed deflection 
geometry, the He atoms exchange almost the same perpendicular 
momentum for different diffraction peaks.
\bibitem{nota2} In fact, an ordinary SCDW propagating on the surface 
(like those proposed 
in Refs. [1,10]) would leave unaltered 
the $(3 \times 3)$ correlation length and the diffracted peaks 
would disappear without broadening.
\end{thebibliography}
\end{document}